\documentclass[conference]{IEEEtran}
\IEEEoverridecommandlockouts
\usepackage{cite}
\usepackage{amsmath,amssymb,amsfonts}
\usepackage{algorithmic}
\usepackage{graphicx}
\usepackage{textcomp}

\usepackage{url}
\usepackage{todonotes}
\usepackage{tabularx}
\usepackage{booktabs} 
\usepackage[T1]{fontenc}
\usepackage{etoolbox}

\usepackage[multiple]{footmisc}
\usepackage{bigfoot}

\usepackage{xcolor}
\def\BibTeX{{\rm B\kern-.05em{\sc i\kern-.025em b}\kern-.08em
    T\kern-.1667em\lower.7ex\hbox{E}\kern-.125emX}}
\begin{document}

\newcommand\gareth[1]{\textbf{\textcolor{red}{GT: #1}}	}

\title{Twitter Dataset for 2022 Russo-Ukrainian Crisis\\
% {\footnotesize \textsuperscript{*}Note: Sub-titles are not captured in Xplore and
% should not be used}
% \thanks{Identify applicable funding agency here. If none, delete this.}
}

% #Hong Kong University of Science and Technology
% Korea Advanced Institute of Technology

\author{
\IEEEauthorblockN{
Ehsan-Ul Haq\IEEEauthorrefmark{1}, 
Gareth Tyson\IEEEauthorrefmark{1}, 
Lik-Hang Lee\IEEEauthorrefmark{2}, 
Tristan Braud\IEEEauthorrefmark{1}, 
and Pan Hui\IEEEauthorrefmark{1}\IEEEauthorrefmark{3}}
\IEEEauthorblockA{\IEEEauthorrefmark{1}Hong Kong University of Science and Technology, Hong Kong SAR}
\IEEEauthorblockA{\IEEEauthorrefmark{2}Korea Advanced Institute of Science and Technology, South Korea}
\IEEEauthorblockA{\IEEEauthorrefmark{3}University of Helsinki, Helsinki, Finland}
Email: euhaq@connect.ust.hk \quad gtyson@ust.hk  \quad likhang.lee@kaist.ac.kr  \quad braudt@ust.hk \quad panhui@cse.ust.hk 
}

\maketitle

\begin{abstract}
Online Social Networks (OSNs) play a significant role in information sharing during a crisis. The data collected during such a crisis can reflect the large scale public opinions and sentiment. In addition, OSN data can also be used to study different campaigns that are employed by various entities to engineer public opinions. Such information sharing campaigns can range from spreading factual information to propaganda and misinformation. We provide a Twitter dataset of the 2022 Russo-Ukrainian conflict. In the first release, we share over 1.6 million tweets shared during the 1st week of the crisis.  
\end{abstract}


\section{Introduction}\label{sec:intro}

The Russo-Ukrainian conflict escalated in February 2022 after Russia recognised two Ukrainian breakaway regions --- the Donetsk People's Republic and the Luhansk People's Republic~\cite{Luhanska10:online}. Following this, the Russian Federation's senate granted the use of military force in those regions on 22\(^{nd}\) February 2022. 
On 24\(^{th}\) February, the Russian government began an invasion of Ukraine, what it referred to as a
\textit{special military operation}~\cite{Putinord95:online}. 

At the time of writing, the conflict is still ongoing. People worldwide have been using social media to share their opinions regarding this conflict. Online Social Networks (OSNs) have been a prominent source of data in studying prior large-scale information discourse during crises and social movements~\cite{ul2020enemy,sacco2015using}, particularly in terms of `information warfare' where such platforms can become the source of propaganda and misinformation~\cite{patrikarakos2017war}. 
To facilitate timely analysis, we are therefore publishing an open dataset relevant to this ongoing crisis. This data can help in studying the political discourse, opinion mining, and (mis)information propagation on Twitter~\cite{haq2020survey}. We will continue to gather data and publish updates every 24 hours.

\section{Dataset Collection}\label{sec:dataset}

We use the Twitter Streaming API\footnote{\url{https://developer.twitter.com/en/docs/twitter-api}} to collect data in real-time.
Note, the Streaming API is the standard method used to collect Twitter data~\cite{chen2020tracking}. 
We use a list of keywords associated with the crisis, as shown in Table~\ref{tab:key_words}. At the time of writing, the crisis is still ongoing. As such, the list of keywords is likely to change over the time with inclusion of the newer words according to the situation. We will incorporate any changes and update the list in future versions. The Twitter streaming API tracks the requested words and provides any tweet that contains any of the words. We will continue the data collection and update the data repository accordingly.

\begin{table}[t]
    \centering
    \caption{List of data collection keywords, including the date they are added to the crawl }
    \begin{tabular}{|c|c|}
    \toprule
        \textbf{Date} & \textbf{Keywords}\\
        \toprule
        27, Feb, 2022 & russia , ukraine , putin , zelensky \\
        
       &  russian , ukrainian , keiv , kyiv \\
       \hline
      1, March, 2022 &  kharkiv \\
      \hline
      3, March, 2022 & khorsan \\
      \hline
      4, March, 2002 & zaporizhzhia, energodar \\
        \bottomrule
    \end{tabular}
    \label{tab:key_words}
\end{table}

\begin{figure}[t]
    \centering
    \includegraphics[width=.4\textwidth]{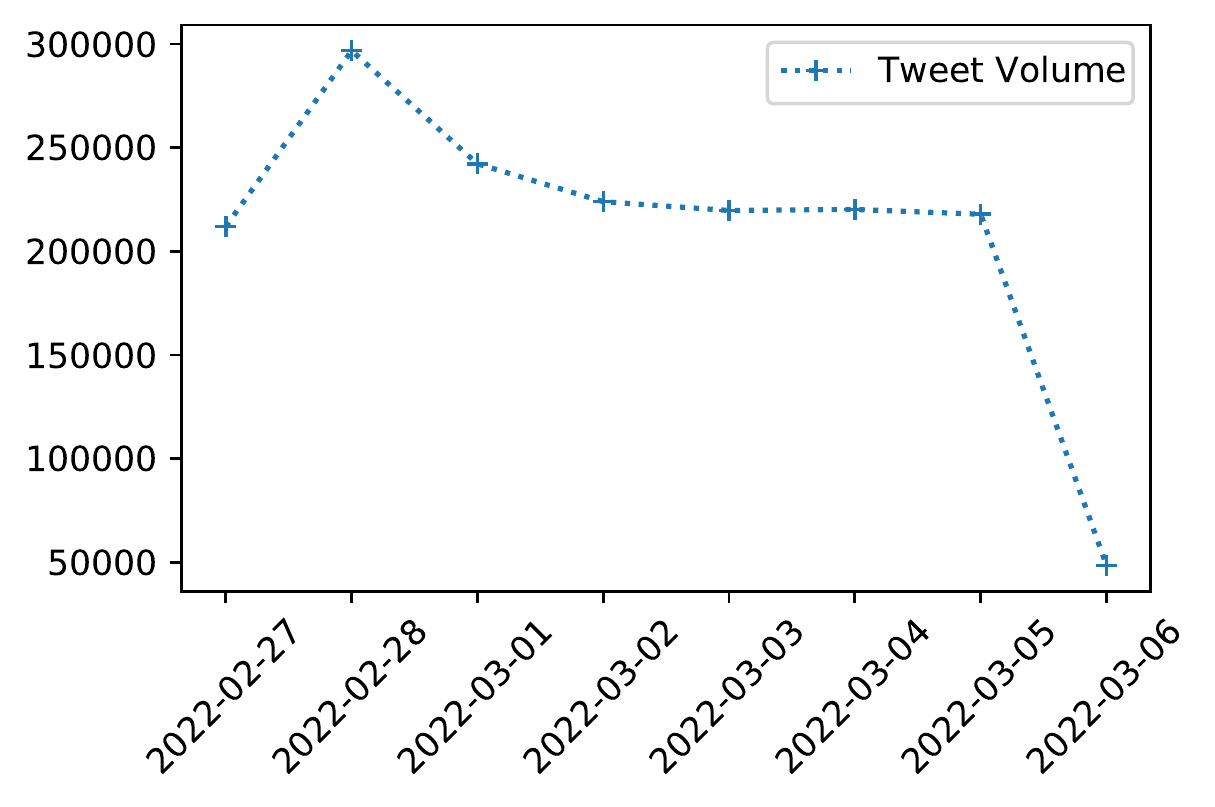}
    \caption{Tweets Daily Frequency}
    \label{fig:frequency}
\end{figure}

\section{Dataset Summary \& Usage}\label{sec:dataset_usage}

\begin{figure}[t]
    \centering
    \includegraphics[width=0.5\textwidth]{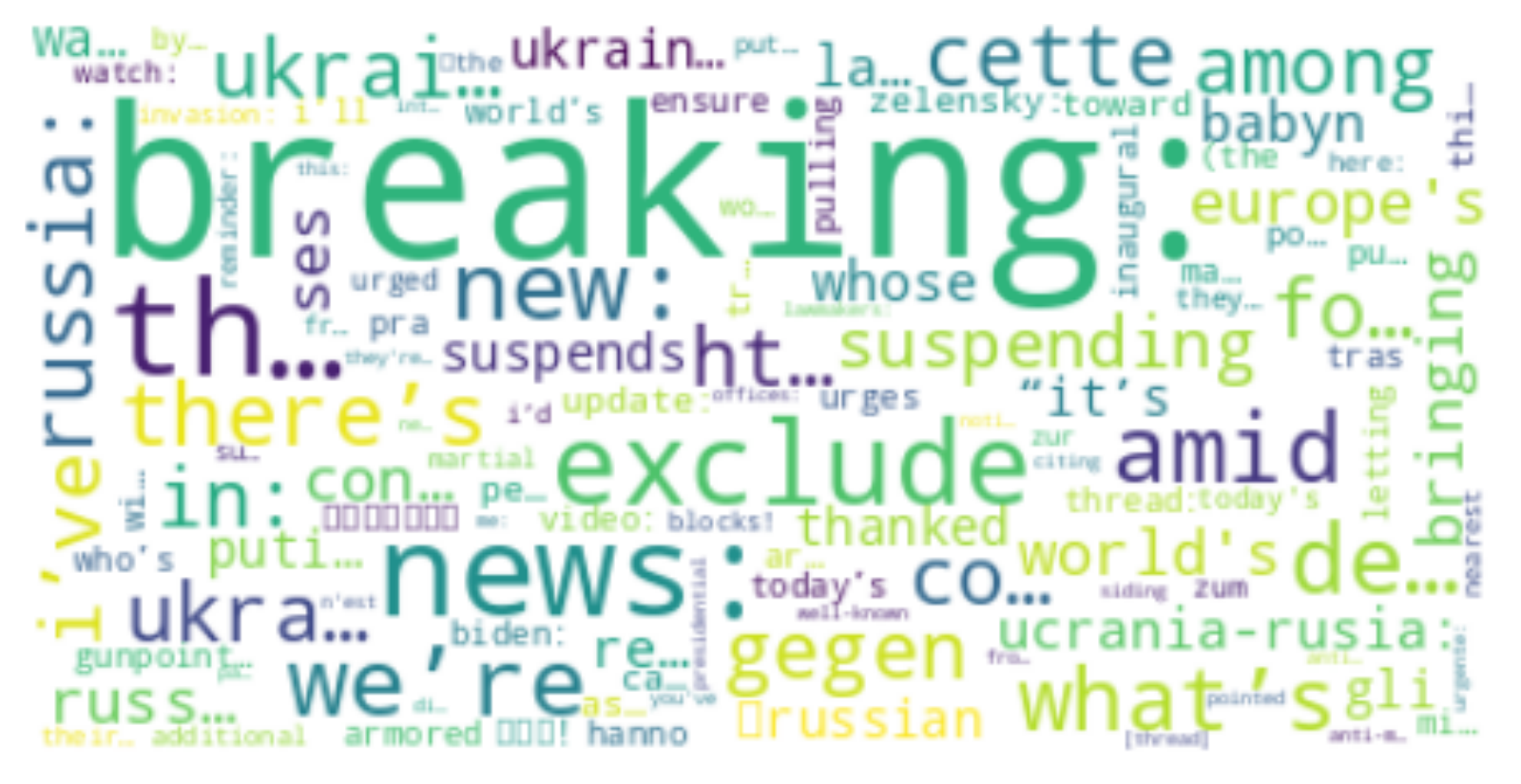}
    \caption{A word cloud of tweets - excluding hashtags and mention}
    \label{fig:word_cloud}
\end{figure}

By the 6\textsuperscript{th} March, we have collected over 1.6 million tweets using the keywords as mentioned in Table~\ref{tab:key_words}.~\footnote{until 6:00 AM, 6\textsuperscript{th} March 2022, UTC Time} We do not apply any language or geo-filters to the data collection. Hence, the dataset contains tweets from different regions in several languages worldwide. The daily volume of tweets is shown in Figure~\ref{fig:frequency}, with an average of about 200K tweets every day. We show the tweet distribution based on the keywords in  Table~\ref{tab:key_words}. Most of the tweets contain the Russia and Ukraine reference, followed by Putin. Zelensky is mentioned less as compared to Putin. 
In addition, we also show the top-10 used hashtags and mentions, across the so far collected data, in Table~\ref{tab:top_10_hts_mts}. In terms of mentions, Zelensky has higher mentions. In addition, most of the mentions are related to western leaders. A word cloud of tweet text is shown in Figure~\ref{fig:word_cloud}. The prominence of words like `breaking', `news', and `suspensions' suggests that most of tweets discuss the latest situational updates. 

There are over 900,000 users in current snapshot of the data. Out of all tweets,  over 1.2 million are retweets. From these tweets, 413,254 unique tweets have been retweeted with an average of three retweets per tweets, while the standard deviation is 12.04.

\begin{table}[t]
\centering
\caption{Number of Tweets containing the keywords}
\begin{tabular}{|ll|}
\toprule
\textbf{Keyword} & \textbf{Tweets}\\
\midrule
putin             & 328186 \\
zelensky          & 86122  \\
russia|russian    & 536464 \\
ukraine|ukrainian & 687321 \\
keiv|Kyiv         & 91142  \\
kharkiv           & 27089  \\
zaporizhzhia      & 8644  \\
\bottomrule
\end{tabular}
\label{tab:tweet_table}
\end{table}

\begin{table}[]
\centering
\caption{Top 10 Hashtags and Mentions}
\begin{tabular}{|l|l|}
\toprule
\textbf{hashtags }        &\textbf{mentions}         \\
\midrule
Ukraine          & ZelenskyyUa     \\
Russia           & NATO            \\
Putin            & POTUS           \\
UkraineRussiaWar & Ukraine         \\
Russian          & UN              \\
Kyiv             & vonderleyen     \\
Ukrainian        & elonmusk        \\
Kharkiv          & KyivIndependent \\
ukraine          & EmmanuelMacron  \\
\bottomrule
\end{tabular}
\label{tab:top_10_hts_mts}
\end{table}
% \gareth{Maybe we could add a few extra stats, e.g. in a table. This could summarise the number of accounts, number of tweets, how many tweets per keyword, nyumber of URLs shared, most popular URL domains etc. We could also add a basic word cloud to give a flavour of what is in the data?}

Data is shared according to Twitter guidelines, governing data usage and sharing~\cite{Develope62:online}. As such, we only share the tweet IDs. We group the tweet IDs by date. This date is according to UTC, as returned by the Twitter API. We plan to update our public repository of data every 24 hours. We share these tweet IDs on a GitHub repository here:

\begin{center}
\texttt{\url{https://github.com/ehsanulhaq1/russo_ukraine_dataset.git}}
\end{center}

The tweet IDs are distributed across multiple files according to the date. Each file contains up to 50,000 tweet IDs.
We recommend open source tools such as Twarc~\cite{DocNowtw4:online}, Tweepy~\cite{Tweepy45:online}, or Hydrator~\cite{DocNowhy97:online} to download the complete tweets using these tweet IDs. 
We will update this repository daily with the latest tweet IDs and relevant statistics. In addition, we will also update the tweets according to the potential warning label as flagged by Twitter.

\bibliographystyle{IEEEtran}
\bibliography{main}

% Generated by IEEEtran.bst, version: 1.14 (2015/08/26)
\begin{thebibliography}{10}
\providecommand{\url}[1]{#1}
\csname url@samestyle\endcsname
\providecommand{\newblock}{\relax}
\providecommand{\bibinfo}[2]{#2}
\providecommand{\BIBentrySTDinterwordspacing}{\spaceskip=0pt\relax}
\providecommand{\BIBentryALTinterwordstretchfactor}{4}
\providecommand{\BIBentryALTinterwordspacing}{\spaceskip=\fontdimen2\font plus
\BIBentryALTinterwordstretchfactor\fontdimen3\font minus
  \fontdimen4\font\relax}
\providecommand{\BIBforeignlanguage}[2]{{%
\expandafter\ifx\csname l@#1\endcsname\relax
\typeout{** WARNING: IEEEtran.bst: No hyphenation pattern has been}%
\typeout{** loaded for the language `#1'. Using the pattern for}%
\typeout{** the default language instead.}%
\else
\language=\csname l@#1\endcsname
\fi
#2}}
\providecommand{\BIBdecl}{\relax}
\BIBdecl

\bibitem{Luhanska10:online}
``Luhansk and donetsk are key to the latest escalation in the ukraine crisis :
  Npr,''
  \url{https://www.npr.org/2022/02/22/1082345068/why-luhansk-and-donetsk-are-key-to-understanding-the-latest-escalation-in-ukrain},
  (Accessed on 03/04/2022).

\bibitem{Putinord95:online}
``Putin orders troops into eastern ukraine on ‘peacekeeping duties’ |
  ukraine | the guardian,''
  \url{https://www.theguardian.com/world/2022/feb/21/ukraine-putin-decide-recognition-breakaway-states-today},
  (Accessed on 03/04/2022).

\bibitem{ul2020enemy}
E.~ul~Haq, T.~Braud, Y.~D. Kwon, and P.~Hui, ``Enemy at the gate: evolution of
  twitter user's polarization during national crisis,'' in \emph{2020 IEEE/ACM
  International Conference on Advances in Social Networks Analysis and Mining
  (ASONAM)}.\hskip 1em plus 0.5em minus 0.4em\relax IEEE, 2020, pp. 212--216.

\bibitem{sacco2015using}
V.~Sacco and D.~Bossio, ``Using social media in the news reportage of war \&
  conflict: Opportunities and challenges,'' \emph{The journal of media
  innovations}, vol.~2, no.~1, pp. 59--76, 2015.

\bibitem{patrikarakos2017war}
D.~Patrikarakos, \emph{War in 140 characters: how social media is reshaping
  conflict in the twenty-first century}.\hskip 1em plus 0.5em minus 0.4em\relax
  Hachette UK, 2017.

\bibitem{haq2020survey}
E.~U. Haq, T.~Braud, Y.~D. Kwon, and P.~Hui, ``A survey on computational
  politics,'' \emph{IEEE Access}, vol.~8, pp. 197\,379--197\,406, 2020.

\bibitem{chen2020tracking}
E.~Chen, K.~Lerman, E.~Ferrara \emph{et~al.}, ``Tracking social media discourse
  about the covid-19 pandemic: Development of a public coronavirus twitter data
  set,'' \emph{JMIR public health and surveillance}, vol.~6, no.~2, p. e19273,
  2020.

\bibitem{Develope62:online}
``Developer agreement and policy – twitter developers | twitter developer
  platform,''
  \url{https://developer.twitter.com/en/developer-terms/agreement-and-policy},
  (Accessed on 03/04/2022).

\bibitem{DocNowtw4:online}
``Docnow/twarc: A command line tool (and python library) for archiving twitter
  json,'' \url{https://github.com/DocNow/twarc}, (Accessed on 03/04/2022).

\bibitem{Tweepy45:online}
``Tweepy,'' \url{https://www.tweepy.org/}, (Accessed on 03/04/2022).

\bibitem{DocNowhy97:online}
D.~Now, ``Docnow/hydrator: Turn tweet ids into twitter json \& csv from your
  desktop!'' \url{https://github.com/DocNow/hydrator}, 2020, (Accessed on
  03/04/2022).

\end{thebibliography}

\end{document}